\documentclass[twocolumn,preprintnumbers,showpacs,pre,floatfix,amsmath,amssymb,notitlepage]{revtex4-1}
\usepackage{epsfig}
\usepackage{subfigure}
\usepackage{amsmath}
\usepackage{color}
\usepackage{amssymb}
\usepackage{setspace}
\usepackage{graphicx}
\usepackage{dcolumn}
\usepackage{bm}
\usepackage{times}
\usepackage{enumerate}
\usepackage{footnote}
\usepackage{cancel}
\input epsf

\newcommand{\appropto}{\mathrel{\vcenter{
  \offinterlineskip\halign{\hfil$##$\cr
    \propto\cr\noalign{\kern2pt}\sim\cr\noalign{\kern-2pt}}}}}

\graphicspath{{figures/}}

\begin{document}

\author{Per Sebastian Skardal}
\email{persebastian.skardal@trincoll.edu} 
\affiliation{Department of Mathematics, Trinity College, Hartford, CT 06106, USA}

\title{Diffusion dynamics and synchronizability of hierarchical products of networks}

\begin{abstract}
The hierarchical product of networks represents a natural tool for building large networks out of two smaller subnetworks: a primary subnetwork and a secondary subnetwork. Here we study the dynamics of diffusion and synchronization processes on hierarchical products. We apply techniques previously used for approximating the eigenvalues of the adjacency matrix to the Laplacian matrix, allowing us to quantify the effects that the primary and secondary subnetworks have on diffusion and synchronization in terms of a coupling parameter that weighs the secondary subnetwork relative to the primary subnetwork. Diffusion processes are separated into two regimes: for small coupling the diffusion rate is determined by the structure of the secondary network, scaling with the coupling parameter, while for large coupling it is determined by the primary network and saturates. Synchronization, on the other hand, is separated into three regimes: for both small and large coupling hierarchical products have poorly synchronization properties, but is optimized at an intermediate value. Moreover, the critical coupling value that optimizes synchronization is shaped by the relative connectivities of the primary and secondary subnetworks, compensating for significant differences between the two subnetworks.
\end{abstract}


\maketitle

\section{Introduction}\label{sec1}

The underlying structures that dictate the patterns of interactions that take place throughout nature and society are often described by complex networks~\cite{Strogatz2001Nature}. Examples of such networks include electrical power grids~\cite{Motter2013NatPhys}, faculty hiring networks~\cite{Clauset2015SciAdv}, gene regulatory networks~\cite{Lee2002Science}, and the structure of academic institutions~\cite{Wang2017ANS}. Many large networks are comprised of smaller subnetwork structures, for example motifs~\cite{Milo2002Science}, communities~\cite{Girvan2002PNAS}, layers~\cite{DeDomenico2013PRX}, self-similar structures~\cite{Guimera2003PRE}, or other subnetwork structures~\cite{Gao2012NatPhys}. In many such cases the properties of the larger network depends on properties of these smaller structures~\cite{Chauhan2009PRE}. Moreover, the collective organization of these smaller subnetwork structures often has a strong effect on the properties of many dynamical processes such as diffusion~\cite{Gomez2013PRL}, synchronization~\cite{Skardal2012PRE} and epidemic spreading~\cite{Liu2005EPL}.

Recently, Barri\`{e}re et al. introduced the hierarchical product~\cite{Barriere2009DAM1,Barriere2009DAM2} as a tool for building a large network using two smaller subnetworks. The hierarchical product is a generalization of the Cartesian product~\cite{Hammack2011}, combining subnetworks in a less regular manner, resulting in a more disordered and heterogeneous structure -- an important characteristic of many real-world networks~\cite{Newman2003SIAM}. Since its introduction, several properties of hierarchical products have been studied, including properties such as radius and diameter, clustering coefficient and degree distribution~\cite{Barriere2016JPA}. Recently, we provided an asymptotic analysis for the full spectrum of the adjacency matrix of the hierarchical product~\cite{Skardal2016PRE}. Here we apply these results in order to study the dynamics that take place on hierarchical products, particularly diffusion and synchronization.

Diffusion and synchronization represent two classical and well-studied classes of dynamical processes on networks. Diffusion has proven to be a particularly versatile tool in network science, identifying structural properties~\cite{Lambiotte2011PRE,Rosvall2014NatComm} and serving as a mathematical model for other relaxation processes~\cite{Grabow2012PRL,Grabow2015PRE,Skardal2016PREb}. Synchronization dynamics are also strongly intertwined with the structures on which they evolve~\cite{Pecora1998PRL}, revealing topological properties of the the underlying networks~\cite{Arenas2006PRL,Pecora2014NatComm}. The long-term dynamics of both diffusion and synchronization dynamics are determined by the eigenvalue spectrum of the network's associated Laplacian matrix. In this work we apply techniques previously used to describe the eigenvalue spectrum of the adjacency matrix of hierarchical products to the Laplacian matrix in order to study the long-term diffusion and synchronization dynamics in this context. Our results allows us to extract the contributions that the two different subnetworks of the hierarchical product have on the long-term the macroscopic dynamics with respect to a coupling parameter that weighs of the secondary subnetwork relative to the the primary subnetwork. In the case of diffusion, two regimes emerge. For small coupling the diffusion rate is slow, increasing along with the coupling, and is determined by the structure of the secondary subnetwork. For large coupling the diffusion rate saturates and is determined by the structure of the primary subnetwork. Thus, a transition in both the dynamics and the contributions from the two subnetworks occurs as the coupling passes through this intermediate range. In the case of synchronization, three regimes emerge. For both small and large coupling the hierarchical product has poor synchronization properties, owing to a large deviation in a large gap between eigenvalues of the Laplacian that results in one subnetwork being weighted significantly more than the other. Synchronization properties are instead optimized at an intermediate, critical coupling value. Interestingly, this critical coupling value highlights the difference in overall connectivity in the primary and secondary subnetworks. Specifically, the critical coupling value is tuned to compensate for either the primary or secondary subnetwork having significantly weaker connectivity than the other. More broadly, the phenomena described in this paper identify the roles that primary and secondary subnetworks play in shaping large-scale dynamics in hierarchical products and which substructure promote vs hinder diffusion and synchronization.

The remainder of this paper is organized as follows. In Sec.~\ref{sec2} we define the hierarchical product and present asymptotic results describing the eigenvalue spectrum of the associated Laplacian matrix. In Sec.~\ref{sec3} we study the long-term behavior of diffusion on hierarchical products. We characterize the behavior of the smallest nontrivial eigenvalue, which dictates the diffusion rate and timescale. Using these results we identify two different regimes where diffusion processes behave differently. In Sec.~\ref{sec4} we study the long-term behavior of synchronization on hierarchical products. Here we characterize both the largest and smallest nontrivial eigenvalues, which determine the synchronizability ratio. This allows us to identify regions of poor synchronization properties for small and large coupling, and optimal synchronization properties at an intermediate, critical coupling value. In Sec.~\ref{sec5} we conclude with a discussion of our results.

\section{The Hierarchical Product}\label{sec2}

\subsection{Coupling Matrices}\label{sec2sub1}

\begin{figure}[t]
\centering
\epsfig{file =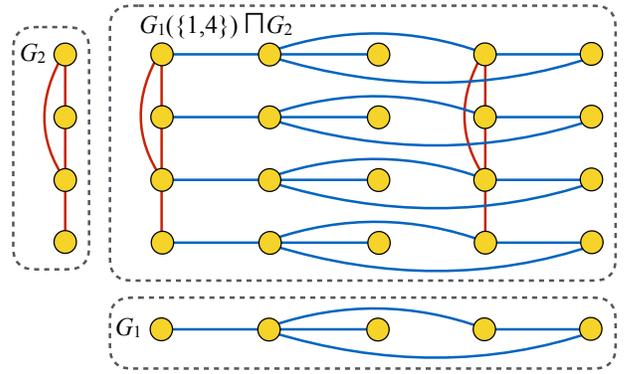, clip =,width=0.95\linewidth }
\caption{(Color online) {\it Hierarchical product}. Illustration of the hierarchical product $G_1(\{1,4\})\sqcap G_2$ of two subgraphs $G_1$ and $G_2$.}
\label{fig1}
\end{figure}

The hierarchical product represents a tool for building a large network from two smaller subnetworks. Here we will consider the hierarchical product of a primary and secondary subnetwork, denoted $G_1$ and $G_2$, respectively, each consisting of $N_1$ and $N_2$ nodes. We will assume that both networks are undirected and binary (or unweighted) so that the adjacency network $A_1$ and $A_2$ have entries $a_{ij}=a_{ji}=1$ if a link exists between nodes $i$ and $j$, and otherwise $a_{ij}=a_{ji}=0$. These properties can be generalized: in a directed network $a_{ij}$ and $a_{ji}$ need not be equal and in a weighted network $a_{ij}$ may take on values aside from zero and one. The {\it hierarchical product} of $G_1$ and $G_2$, denoted $G_1(U)\sqcap G_2$, is a network of $N=N_1\cdot N_2$ nodes that consists of $N_2$ copies of $G_1$ that are each connected to one another through the nodes in {\it the root set} $U$ via the topology of $G_2$. In Fig.~\ref{fig1} the hierarchical product is illustrated using an example with subnetworks $G_1$ and $G_2$ with $N_1=5$ and $N_2=4$ nodes, respectively. As a root set we use $U=\{1,4\}$, indicating that the four copies of $G_1$ are each connected by $G_2$ through nodes $1$ and $4$, so that in total $G_1(U)\sqcap G_2$ has $N=20$ nodes. The hierarchical product can be further generalized to include the product of an arbitrary number of subnetworks~\cite{Barriere2009DAM2}, however such cases can be defined recursively, so we focus on hierarchical products of two subnetworks. 

The goal of this work is to determine the effects that the two different subnetworks that make up a hierarchical product have on the long-term dynamics of diffusion and synchronization processes. To this end, we introduce a coupling parameter, denoted $\alpha$, that weighs the contribution of the secondary subnetwork $G_2$ in comparison to the primary subnetwork $G_1$. We incorporate this coupling parameter into the adjacency matrix of the hierarchical product, which is given by
 \begin{align}
A_\alpha &= I_2 \otimes A_1 + \alpha A_2 \otimes D_1,\label{eq:01}
\end{align}
where $\otimes$ denotes the Kronecker product, $I_2$ is the $N_2\times N_2$ identity matrix, and $D_1$ is the $N_1\times N_1$ diagonal matrix whose $i^{\text{th}}$ diagonal entry is equal to one if vertex $i$ is in the root set $U$ and zero otherwise otherwise. Thus, $\alpha<1$ and $\alpha>1$ correspond to the links of the secondary subnetwork being weighted weaker and stronger, respectively, than the links of the primary subnetwork. While the spectral properties of Eq.~(\ref{eq:01}) were studied in Ref.~\cite{Skardal2016PRE}, in this work we are interested in the Laplacian matrix due to the role it plays in dynamical processes, specifically diffusion and synchronization. For a network with adjacency matrix $A$, the Laplacian $L$ has entries defined $l_{ij}=\delta_{ij}\left(\sum_{j=1}^N a_{ij}\right)-a_{ij}=\delta_{ij}k_i-a_{ij}$, where $k_i$ denotes the nodal degree. In the case of a hierarchical product with adjacency matrix as in Eq.~(\ref{eq:01}), the Laplacian matrix is given by
 \begin{align}
L_\alpha &= I_2 \otimes L_1 + \alpha L_2 \otimes D_1,\label{eq:02}
\end{align}
where $L_1$ and $L_2$ are the Laplacian matrices of networks $G_1$ and $G_2$, respectively. The eigenvalue spectrum of the Laplacian matrix $L$ of any connected and undirected network has several important properties. First, since every row sums to zero there exists a trivial eigenvalue $\lambda_1=0$ whose associated eigenvector is constant, $\bm{w}^1\propto\bm{1}=[1,\dots,1]^T$. All other eigenvalues are real and positive, so they can be ordered $0=\lambda_1<\lambda_2\le\dots\le\lambda_N$. Finally, the eigenvectors are orthogonal and can therefore be normalized to form an orthonormal basis for $\mathbb{R}^N$.

\subsection{Eigenvalues}\label{sec2sub2}

\begin{figure}[t]
\centering
\epsfig{file =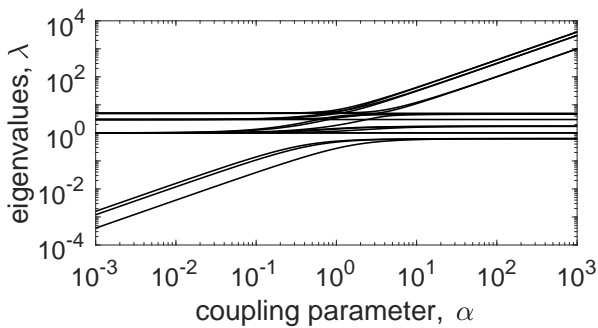, clip =,width=0.98\linewidth }
\caption{(Color online) {\it Laplacian eigenvalues}. The full spectrum of non-trivial eigenvalues for the Laplacian matrix $L_\alpha$ as a function of the coupling parameter $\alpha$ for the hierarchical product illustrated in Fig.~\ref{fig1}.}
\label{fig2}
\end{figure}

The long-term dynamics of both diffusion and synchronization processes depend on the eigenvalues of the associated Laplacian matrix. Thus, for a full understanding of the long-term dynamics on the hierarchical product, we require a characterization of the eigenvalues associated with Eq.~(\ref{eq:02}). In a previous publication we provided an asymptotic analysis for the eigenvalue spectrum of the adjacency matrix of the hierarchical product, i.e., Eq.~(\ref{eq:01})~\cite{Skardal2016PRE}. Here we apply the same methodology to the case of the Laplacian. Our goal is to classify the eigenvalues of $L_\alpha$ in terms of the eigenvalues and eigenvectors of the Laplacian of its subnetworks $L_1$ and $L_2$, as well as the coupling parameter $\alpha$ and the root set encapsulated in the matrix $D_1$. We will denote the eigenvalues of $L_1$ and $L_2$ as $\nu_i$ and $\mu_i$, respectively, and the associated eigenvectors $\bm{v}^i$ and $\bm{u}^i$, respectively. We seek the eigenvalues, denote $\lambda$, of $L_\alpha$. Before classifying them, we illustrate the general behavior of the eigenvalues of the hierarchical product as a function of the coupling parameter $\alpha$ in Fig~\ref{fig2}, using the network illustrated in Fig.~\ref{fig1} as an example. Broadly speaking, these eigenvalues split into two groups for both small and large $\alpha$. For small $\alpha$ one group of eigenvalues are themselves small, scaling approximately as $\alpha$ and the other remains approximately constant, on the order of one. For large $\alpha$ one group of eigenvalues also remains approximately constant, on the order of one, but the other group is itself large, also scaling approximately as $\alpha$. When $\alpha$ is itself on the order of one these groups coalesce in a complicated arrangement.

The classification of the eigenvalues $\lambda$ then begins with the analysis of a new set of matrices. Specifically, $\lambda$ is an eigenvalue of $L_\alpha$ if and only if it is also an eigenvalue of one of the matrices given by
\begin{align}
L_\alpha(\mu_i)=L_1+\alpha\mu_i D_1,\label{eq:03}
\end{align}
where $L_\alpha(\mu_i)$ is one of the $N_2$ possible $N_1\times N_1$ matrix constructed via a linear combination of $L_1$ and $D_1$, where $D_1$ is scaled by one of the $N_2$ eigenvalues $\mu_i$ of $L_2$~\cite{Barriere2009DAM1}. In total, there are $N_2$ such matrices $L_\alpha(\mu_i)$, each of which have $N_1$ eigenvalues, resulting in the full spectrum of $N_1\cdot N_2$ eigenvalues of $L_\alpha$. Thus, the eigenvalue problem of $L_\alpha$ is reduced to the set of smaller eigenvalue problems for the collection of $L_\alpha(\mu_i)$.

While the collection of eigenvalues of Eq.~(\ref{eq:03}) can be found perturbatively as in Ref.~\cite{Skardal2016PRE}, a specific subset deserves particular attention here. Recall that the Laplacian of a connected network has precisely one zero eigenvalue. Thus, $L_2$ has one zero eigenvalue $\mu_1=0$. Inserting this into Eq.~(\ref{eq:03}) yields, simply,
\begin{align}
L_\alpha(\mu_1=0)=L_1.\label{eq:04}
\end{align}
Eq.~(\ref{eq:04}) implies then that precisely $N_1$ of the eigenvalues of $L_\alpha$ are given by the eigenvalues $\nu_i$ of $L_1$, and that these eigenvalues remain constant regardless of the value of $\alpha$. Exactly one of these eigenvalues corresponds to the zero eigenvalue $\nu_1=0$ of $L_1$, with the other $N_1-1$ being finite.

As for the remaining $N_1(N_2-1)$ eigenvalues of $L_\alpha$, we apply a perturbative analysis for the limits of small and large coupling. Here we present the results and leave the details for the interested reader in Appendix~\ref{app:A}. In the case of small coupling, i.e., $\alpha\ll1$ the eigenvalues are given, to first order in $\alpha$, by
\begin{align}
\lambda(\alpha)=\nu_j+\alpha\mu_i\bm{v}^{jT}D_1\bm{v}^j,\label{eq:05}
\end{align}
which represents the contributions of the two subnetworks via $\nu_j$ and $\mu_i$ ($j=1,\dots,N_1$, $i=2,\dots,N_2$) to the eigenvalue spectrum. It should be noted that using $\mu_1=0$ in Eq.~(\ref{eq:05}) recovers the constant eigenvalues from Eq.~(\ref{eq:04}), however we consider this a separate case since the eigenvalues from Eq.~(\ref{eq:04}) are exact, while those in Eq.~(\ref{eq:05}) are approximations.

\begin{figure}[t]
\centering
\epsfig{file =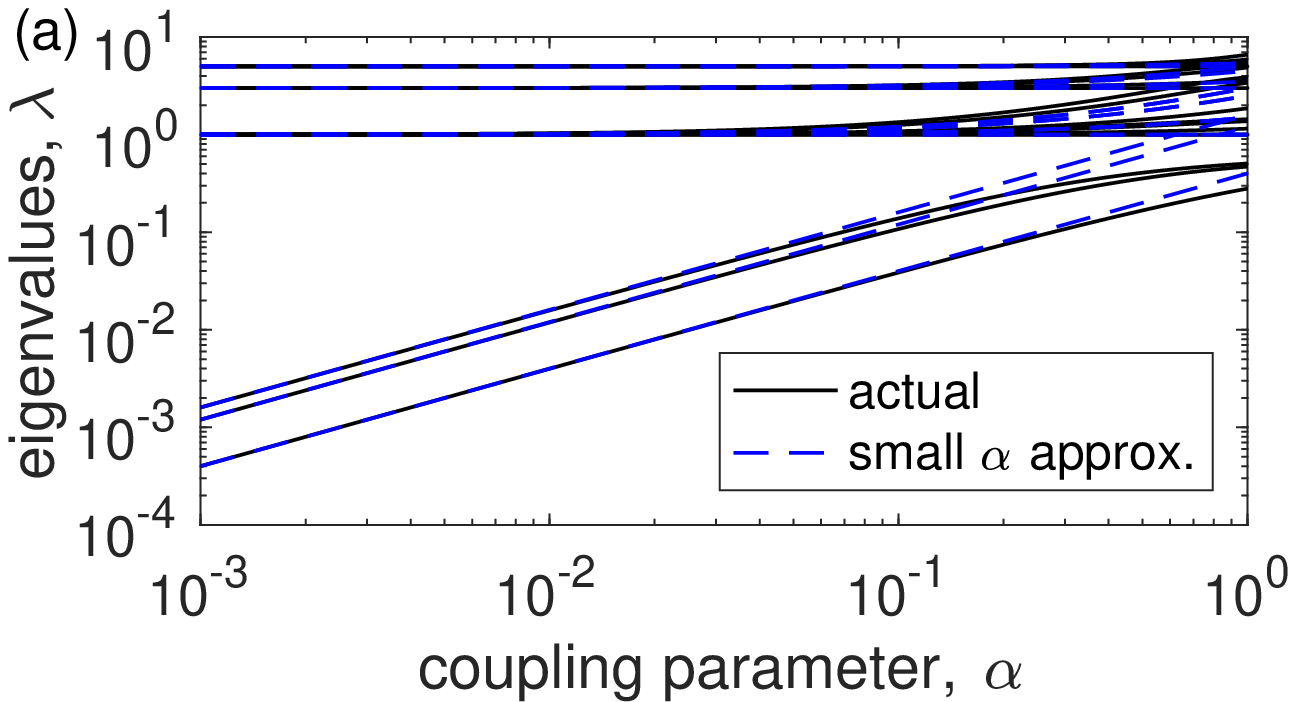, clip =,width=0.98\linewidth }
\epsfig{file =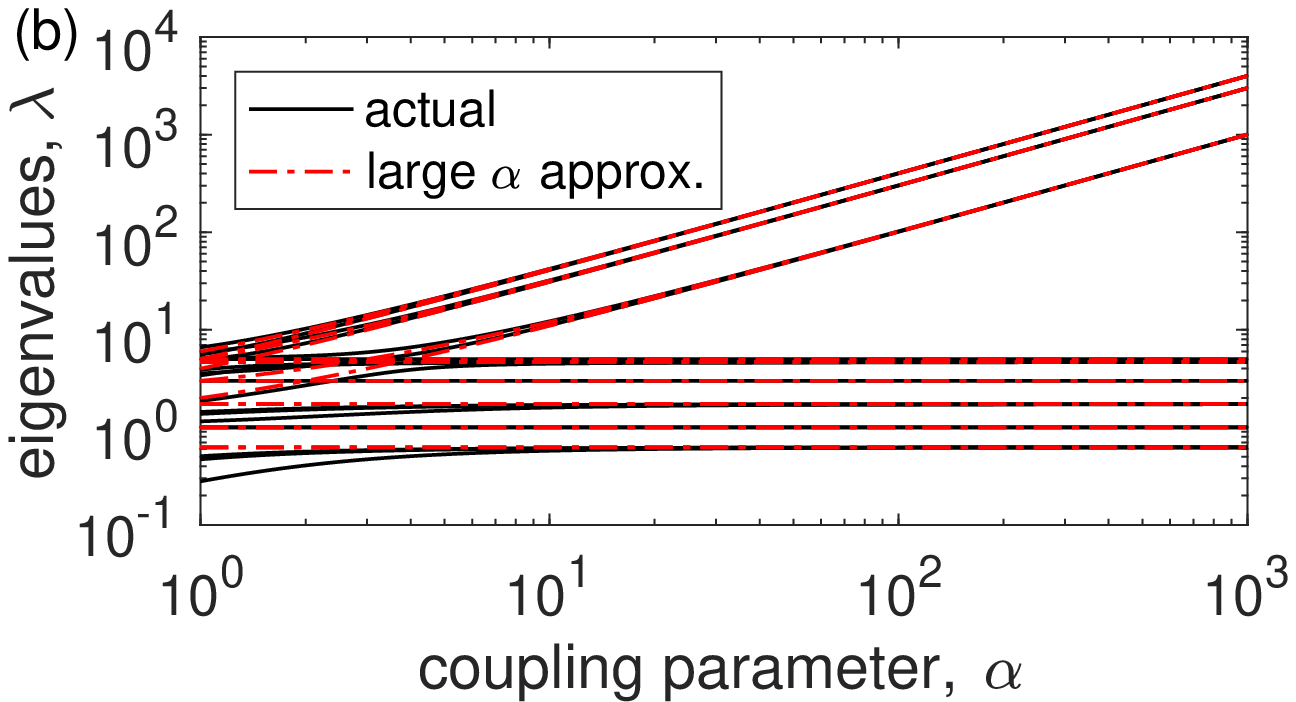, clip =,width=0.98\linewidth }
\caption{(Color online) {\it Laplacian eigenvalues}. Approximate (dashed blue and dot-dashed red) and actual (solid black) eigenvalues for the Laplacian matrix $L_\alpha$ for the hierarchical product illustrated in Fig.~\ref{fig1} for (a) small and (b) large $\alpha$.}
\label{fig3}
\end{figure}

In the limit of large coupling, i.e., $\alpha\gg1$, the analysis becomes more complicated with the $N_1(N_2-1)$ eigenvalues splitting into two distinct groups due to the degeneracy of $D_1$. The first group yields $n(N_2-1)$ eigenvalues, where $n$ is the size of the root set and the number of nonzero entries of $D_1$, and are given by
\begin{align}
\lambda(\alpha)=\alpha\mu_i+\nu_j^{\cancel{0}},\label{eq:06}
\end{align}
where $\nu_j^{\cancel{0}}$ is an eigenvalues of the $n\times n$ matrix $L_1^{\cancel{0}}$ constructed by removing all rows and columns of $L_1$ corresponding to zero diagonal entries of $D_1$. The remaining $(N_1-n)(N_2-1)$ eigenvalues are then given by
\begin{align}
\lambda(\alpha)=\nu_j^{0},\label{eq:07}
\end{align}
where $\nu_j^{0}$ is an eigenvalues of the $(N_1-n)\times(N_1-n)$ matrix $L_1^{0}$ constructed by removing all rows and columns of $L_1$ corresponding to nonzero diagonal entries of $D_1$. In Figs.~\ref{fig3}(a) and (b) we compare the analytical approximations (dashed blue and dot-dashed red curves) for the eigenvalues of $L_\alpha$ to the actual values (solid black) for the example hierarchical product illustrated in Fig.~\ref{fig1} for small and large $\alpha$. In both cases the approximations accurately describe the behavior of the eigenvalues for sufficiently small and large $\alpha$, with the approximations breaking down when $\alpha$ is approximately of order one.

\section{Diffusion}\label{sec3}

We now turn our attention to the long-term dynamics of diffusion process on hierarchical products. Given an adjacency matrix with entries $a_{ij}$, diffusion is governed by the following equations
\begin{align}
\dot{x}_i=\sum_{j=1}^Na_{ij}(x_j-x_i),\label{eq:08}
\end{align}
which can be rewritten in vector form as
\begin{align}
\dot{\bm{x}}=-L\bm{x},\label{eq:09}
\end{align}
where $\bm{x}$ is the state vector of the process and $L$ is the Laplacian. We note here that we focus on the specific case of diffusion related to heat transfer and relaxation dynamics, in which case we use the combinatorial Laplacian $L=D-A$, where $D=\text{diag}(n_1\dots,k_N)$. In the case of diffusion related to a random walk processes, the symmetric or asymmetric versions of the normalized Laplacian, $L = I-D^{-1/2}AD^{-1/2}$ or $L=I-D^{-1}A$, may be used. We note that in either case the methodology for approximating eigenvalues may be preserved, but in the asymmetric case the emergence of complex eigenvalues may require more care when applying these results.

Assuming that the underlying network is connected, the dynamics of Eqs.~(\ref{eq:08}) and (\ref{eq:09}) relax to the steady state $x_1=\cdots=x_N=x_{\infty}$ in the limit $t\to\infty$. This relaxation is exponential, specifically with
\begin{align}
\|\bm{x}(t)-\bm{x}_\infty\|\appropto e^{-\lambda_2 t},\label{eq:10}
\end{align}
i.e., the rate of diffusion is given by smallest nontrivial eigenvalue $\lambda_2$ and the timescale of diffusion is given by its inverse $\lambda_2^{-1}$. 
Therefore, we seek specifically the smallest nontrivial eigenvalue $\lambda_2$ from our approximations above. In the small coupling regime, $\alpha\ll1$, the full set of eigenvalues is given by the collection of $N_1$ eigenvalues of $L_1$ along with the $N_1(N_2-1)$ eigenvalues in Eq.~(\ref{eq:05}). Since the nontrivial eigenvalues of $L_1$ are all of order one, the smallest nontrivial eigenvalue is given by using $\nu_1=0$ and $\mu_2$ in Eq.~(\ref{eq:05}), resulting in 
\begin{align}
\lambda_2(\alpha) = \alpha\mu_2\bm{v}^{1T}D_1\bm{v}^1=\frac{\alpha\mu_2 n}{N_1},\label{eq:11}
\end{align}
where we have used that $\bm{v}_1=\bm{1}/\sqrt{N_1}$. In the large coupling regime, $\alpha\gg1$ the full set of eigenvalues is given, again, by the collection of $N_1$ eigenvalues of $L_1$, along with those given in Eqs.~(\ref{eq:06}) and (\ref{eq:07}). Inspecting all possible combinations, the smallest nontrivial eigenvalue is then given by either the smallest nontrivial eigevnalue of $L_1$, or the smallest eigenvalue of $L^0$, i.e., 
\begin{align}
\lambda_2(\alpha) = \text{min}(\nu_2,\nu_1^0).\label{eq:12}
\end{align}
In general, it is impossible to determine a priori which eigenvalue in Eq.~(\ref{eq:12}) in smallest; as we shall see below, different combinations of different subnetwork structures yield different outcomes.

\begin{figure}[t]
\centering
\epsfig{file =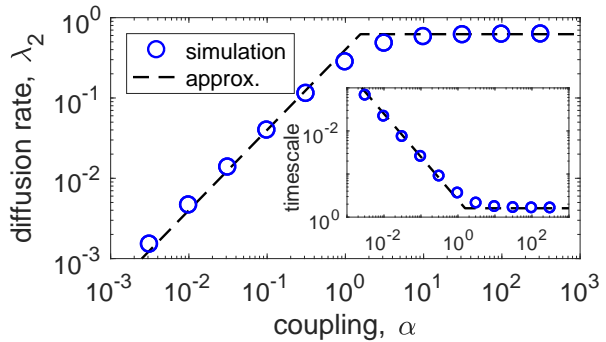, clip =,width=0.98\linewidth }
\caption{(Color online) {\it Diffusion dynamics}. The diffusion rate computed directly from simulations (blue circles) and our approximations for $\lambda_2$ (dashed black) as a function of $\alpha$ for the hierarchical product illustrated in Fig.~\ref{fig1}. Inset: diffusion timescale.}
\label{fig4}
\end{figure}

We first compare our the predictions of our approximations to direct simulation results. In Fig.~\ref{fig4} we plot our theoretical prediction of the diffusion rate (dashed black), i.e., Eqs.~(\ref{eq:11}) and (\ref{eq:12}) for the small and large coupling regimes, respectively, to the diffusion rate observed from simulations (blue circles) on the hierarchical product illustrated in Fig.~\ref{fig1}. Simulated results are computed by fitting the simulations after a significant transient to an exponential. The diffusion timescale is plotted in the inset. The two different dynamical behaviors, corresponding to small and large coupling, are observed and are well captured by the predictions. Specifically, in the small coupling regime the diffusion rate is very slow, and scales with the coupling parameter $\alpha$, which can be observed directly from Eq.~(\ref{eq:11}). Moreover, Eq.~(\ref{eq:11}) reveals that the long-time diffusion dynamics in the small coupling regime is completely determined by the structure of the secondary subnetwork via the eigenvalue $\mu_2$ and the size of the root set via the fraction $n/N_1$. 

In contrast to the small coupling regime, in the large coupling regime, the diffusion rate saturates to the order one value given in Eq.~(\ref{eq:12}). Moreover, Eq.~(\ref{eq:12}) reveals that in this regime the long-time diffusion dynamics in the large coupling regime are completely determined by the structure of the primary subnetwork via the eigenvalue $\nu_2$, and possibly in combination with the root set via the eigenvalue $\nu_1^0$. This highlights a transition between small and large coupling from two perspectives. First, the rate of diffusion itself is increasing, scaling with $\alpha$, for small coupling, and saturates to a constant value for large $\alpha$. Second, the role that the components of the hierarchical product play in this behavior changes; for small $\alpha$ dynamics are dictated by the secondary subnetwork and for large $\alpha$ dynamics are dictated by the primary subnetwork. 

\begin{figure}[t]
\centering
\epsfig{file =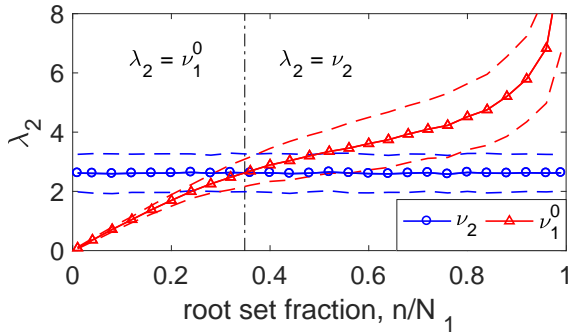, clip =,width=0.98\linewidth }
\caption{(Color online) {\it Large coupling: diffusion rate and root set fraction I}. For ER networks of size $N_1=100$ with link probability $p=0.1$, the quantities $\nu_2$ (blue circles) and $\nu_1^0$ (red triangles) as a function of the root set fraction $n/N_1$. Results represent an average with standard deviation indicated by dashed curves. The transition from $\lambda_2=\nu_1^0$ to $\lambda_2=\nu_2$ occurs at $n/N_1\approx0.3487$.}
\label{fig5}
\end{figure}

Next we investigate in more detail the large coupling regime specifically the determination of $\lambda_2$ as $\nu_2$ or $\nu_1^0$ in Eq.~(\ref{eq:12}). In both cases we note that the structure of the secondary subnetwork is irrelevant -- only the primary subnetwork and the root set determine these quantities. However, in which cases $\lambda_2$ is determined by with choice in unclear. To better understand these quantities, we compare them in Fig.~\ref{fig5}, plotting $\nu_2$ (blue circles) and $\nu_1^0$ (red triangles) computed from a collection of $1000$ realizations of Erd\H{o}s-R\'{e}nyi (ER) networks~\cite{Erdos1960} of size $N=100$ constructed with link probability $p=0.1$ as a function of different root sets fractions $n/N_1$, where nodes in the root sets are randomly chosen. Results represent the average over the $1000$ networks, with the standard deviation denoted by dashed curves. As the root set fraction increases the value $\nu_2$ remains constant (as should be expected for a set network model) and $\nu_1^0$ increases from zero. Thus, for small $n/N_1$ we have $\nu_1^0<\nu_2$, indicating that $\lambda_2=\nu_1^0$, but for large enough $n/N_1$ we have $\nu_2<\nu_1^0$, indicating that $\lambda_2=\nu_2$. For the network model chosen here we find that this transition occurs at $n/N_1\approx0.3487$, which is illustrated with the vertical dot-dashed line. Physically, this suggests that for a small enough root set the diffusion rate is determined by a combination of the structure of the primary subnetwork and the root set itself, but for a large enough root set the diffusion rate is determined solely by the primary subnetwork.

\begin{figure}[t]
\centering
\epsfig{file =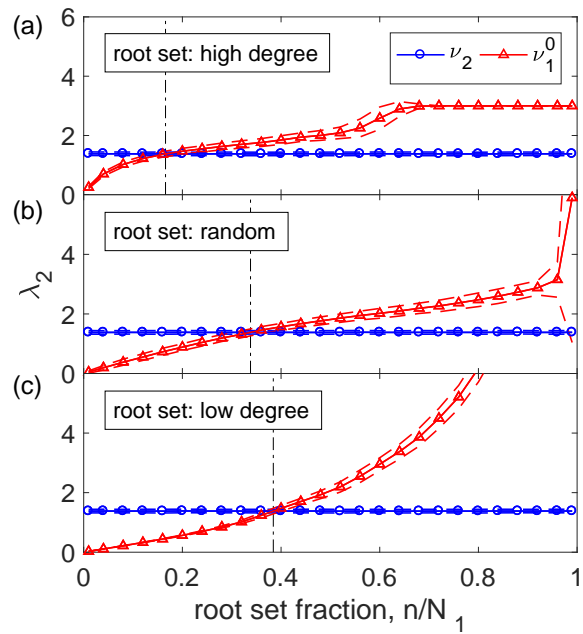, clip =,width=0.98\linewidth }
\caption{(Color online) {\it Large coupling: diffusion rate and root set fraction II}. For BA networks of size $N_1=100$ with minimum degree $k_0=3$, the quantities $\nu_2$ (blue circles) and $\nu_1^0$ (red triangles) as a function of the root set fraction $n/N_1$ for root sets chosen (a) with the highest degree nodes (b) randomly, and (c) with the lowest degree nodes. Results represent an average with standard deviation indicated by dashed curves. For the three cases the transition from $\lambda_2=\nu_1^0$ to $\lambda_2=\nu_2$ occurs at $n/N_1\approx0.1630$, $0.3375$, and $0.3843$, respectively.}
\label{fig6}
\end{figure}

Given that the size of the root set plays a role in whether $\lambda_2$ is given by $\nu_2$ or $\nu_1^0$, it is natural to ask whether the particular locations of the root set also plays a role. In other words, does the behavior of $\nu_2$ and $\nu_1^0$ depend significantly on which nodes belong to the root set, in addition to the size of it? The ER model used in Fig.~\ref{fig5} yields relatively homogeneous networks where nodes have by-and-large very similar structural properties. To investigate this new question we then use the Barabasi Albert (BA) model~\cite{Barabasi1999Science}, which yields much more heterogeneous networks. Specifically, we consider BA networks of size $N=100$ with minimum degree $k_0=3$, but choose the root set in three different ways. In Fig.~\ref{fig6} we plot the average behavior of $\nu_2$ (blue circles) and $\nu_1^0$ (red triangles) as a function of the root set fraction $n/N_1$, choosing the root set to contain (a) the highest degree nodes in the network, (b) randomly selected nodes, and (c) the lowest degree nodes in the network. The generic behavior is similar in the sense that $\nu_2$ remains constant and $\nu_1^0$ increases with $n/N_1$. However, the critical root set fraction at which the transition from $\lambda_2=\nu_1^0$ to $\lambda_2=\nu_2$ occurs is different. Specifically, when the root set consists of the highest degree nodes in the network this transition occurs quite early, at $n/N_1\approx 0.1630$. Conversely, when the root set consists of the lowest degree nodes in the network this transition occurs quite late, at $n/N_1\approx 0.3843$. When the root set consists of randomly chosen nodes this transition occurs in between, at $n/N_1\approx 0.3375$. Thus, when the root set consists of lower degree nodes, it plays a role in determining the diffusion rate for a larger range of root set fractions than when it consists of higher degree nodes.

\section{Synchronization}\label{sec4}

Next we turn to synchronization on hierarchical products. Specifically, we consider synchronization of identical, chaotic dynamical systems, whose dynamics are governed by
\begin{align}
\dot{\bm{x}}_i= \bm{F}(\bm{x}_i)+K\sum_{j=1}^Na_{ij}[\bm{H}(\bm{x}_j)-\bm{H}(\bm{x}_i)],\label{eq:13}
\end{align}
where $\bm{x}_i$ is the state vector of node $i$, $\bm{F}(\bm{x})$ is the (assumed chaotic) vector field describing the internal dynamics of each node, $K$ is the global coupling strength, and $\bm{H}(\bm{x})$ is the coupling function. The dynamics of Eq.~(\ref{eq:13}) are typically treated by studying the stability of the synchronized state $\bm{x}_1(t)=\cdots=\bm{x}_N(t)$, which can be determined using the Master Stability Function (MSF) approach~\cite{Pecora1998PRL}. In particular, the synchronized state is linear stable if all the nontrivial eigenvalues of the Laplacian matrix scaled by the coupling strength $K$ fall within an appropriately defined region of stability. (For the sake of brevity, we forgo a discussion of further technical details and refer the interested reader to the original work in Ref.~\cite{Pecora1998PRL}.) While the region of stability depends on the particular dynamical system $\bm{F}$ and the coupling function $\bm{H}$ in Eq.~(\ref{eq:13}), in many cases the region of stability is a finite interval, denoted $[\gamma_l,\gamma_u]$~\cite{Huang2009PRE}. Synchronization can then be achieved if a coupling $K$ can be chosen such that $K\lambda_i$ for $i=2,\dots,N$ fall within the interval. This is true if and only if the eigenvalues satisfy
\begin{align}
R~\dot{=}~\frac{\lambda_N}{\lambda_2}<\frac{\gamma_u}{\gamma_l},\label{eq:14}
\end{align}
where $R$ is the synchronizability ratio of the network. In particular, the smaller the synchronizability ratio $R$ a given network has, the more synchronizable it is. 

The synchronizability of a given hierarchical product then requires both the smallest and largest nontrivial eigenvalues, $\lambda_2$ and $\lambda_N$. Since we characterized in detail the smallest nontrivial eigenvalue in the previous section, we now turn to the largest. In the small coupling regime we refer back to Eq.~(\ref{eq:05}). Note that these eigenvalue are always larger than the corresponding constant eigenvalues of $L_1$ since $\alpha\mu_i\bm{v}^{jT}D_1\bm{v}^j>0$. Thus, the maximum is obtained by choosing $j=N_1$ and $i=N_2$, yielding
\begin{align}
\lambda_N(\alpha)=\nu_{N_1}+\alpha\mu_{N_2}\bm{v}^{N_1T}D_1\bm{v}^{N_1}.\label{eq:15}
\end{align}
In the large coupling regime, we find that the largest eigenvalue is given by Eq.~(\ref{eq:06}), using $i=N_2$ and $j=n$, yielding
\begin{align}
\lambda_N(\alpha)=\alpha\mu_{N_2}+\nu_n^{\cancel{0}}.\label{eq:16}
\end{align}

\begin{figure}[t]
\centering
\epsfig{file =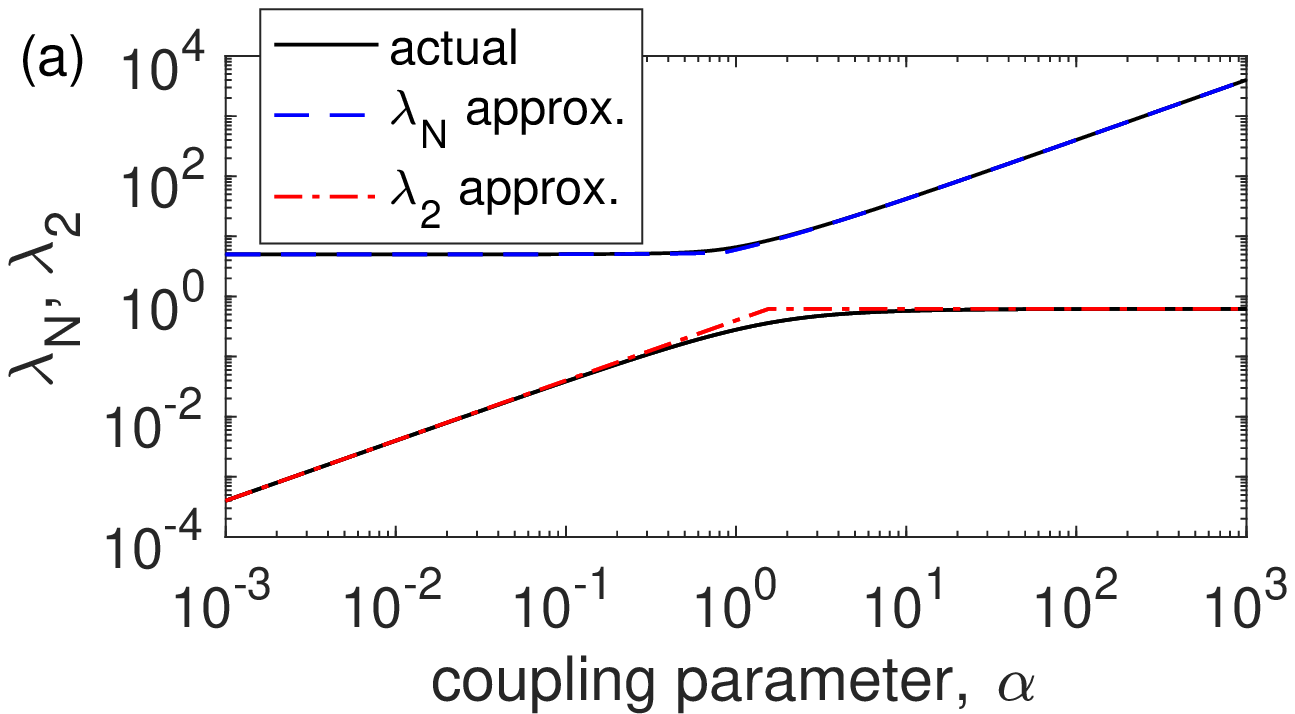, clip =,width=0.98\linewidth }
\epsfig{file =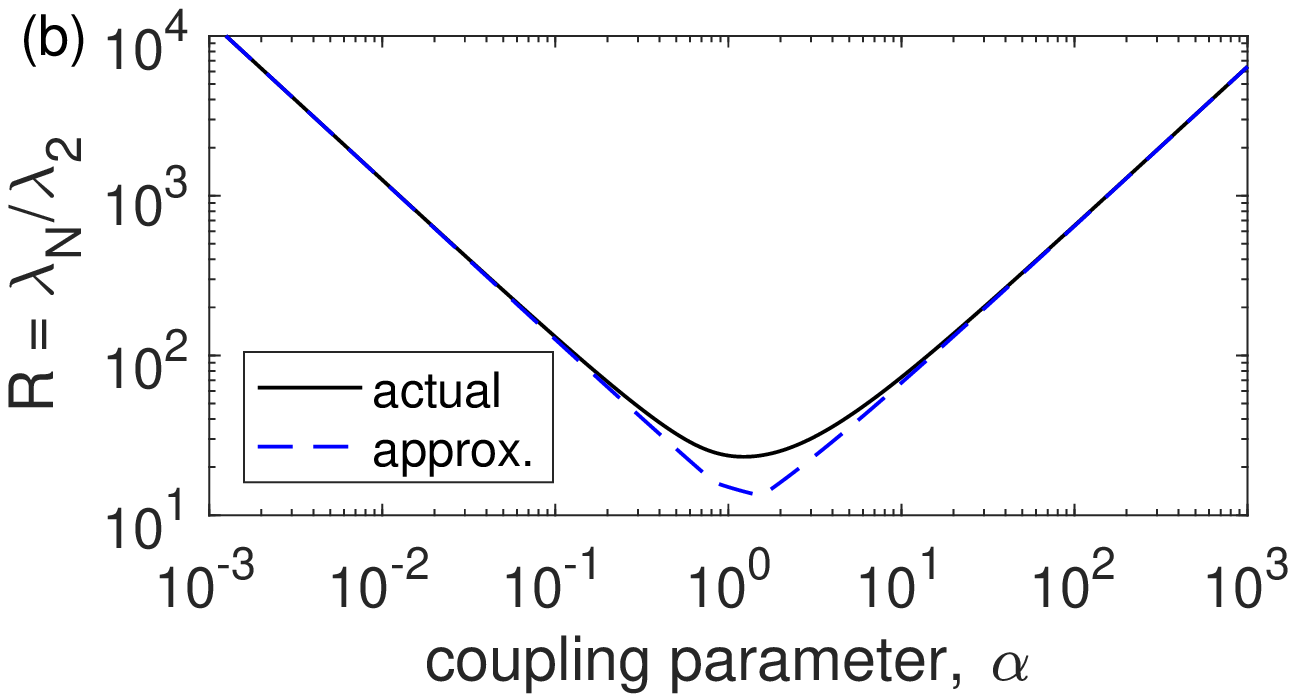, clip =,width=0.98\linewidth }
\caption{(Color online) {\it Synchronizability}. (a) Actual (solid black) and approximate (dashed blue and dot-dashed red, respectively) eigenvalues $\lambda_N$ and $\lambda_2$ as a function of  coupling $\alpha$ for the hierarchical product illustrated in Fig.~\ref{fig1}. (b) Actual (solid black) and approximate (dashed blue) synchronizability ratio $R=\lambda_N/\lambda_2$ as a function of coupling $\alpha$.}
\label{fig7}
\end{figure}

In Figs.~\ref{fig7}(a) and (b) we demonstrate how the synchronizability of the hierarchical product illustrated in Fig.~\ref{fig1} behaves as a function of the coupling parameters, first plotting the separate behaviors of the actual (solid black) approximate (dashed blue and dot-dashed red) values of $\lambda_N$ and $\lambda_2$ in panel (a), then in panel (b) the actual (solid black) and approximate (dashed blue) synchronizability ratio $R=\lambda_n/\lambda_2$. Specifically, we see in panel (b) that for both very large and very small $\alpha$ the synchronizability ratio is large, indicating that the hierarchical product has poor synchronization properties. This is due to the large gap between $\lambda_2$ and $\lambda_N$ which can be observed directly in panel (a), and can be physically attributed to one of the two subnetworks being weighted much heavier than the other. Instead, the hierarchical product displays the best synchronizability ratio for intermediate values of $\alpha$, suggesting that hierarchical products have have the best synchronization properties when the two subnetworks are weighted roughly equally.

\begin{figure}[t]
\centering
\epsfig{file =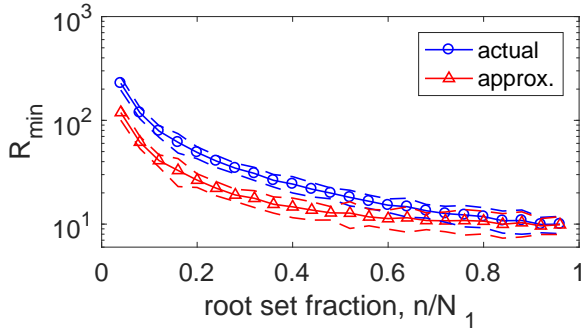, clip =,width=0.98\linewidth }
\caption{(Color online) {\it Optimal synchronizability}. The actual (blue circles) and approximated (red triangles) optimal synchronizability ratio $R_{\text{min}}$ achievable as a function of the root set fraction $n/N_1$. Results represent an average over $1000$ hierarchical products constructed using ER networks of sizes $N_1=100$ and $N_2=20$ with link probabilities $p=0.1$ and $0.5$..}
\label{fig8}
\end{figure}

Next we investigate the role that the two different subnetworks and the root set play in determining the sychronizability of the hierarchical product. First we consider the synchronizability itself -- specifically the optimal (minimal) synchronizability attainable for a given hierarchical product. We find that this quantity depends primarily on the size of the root set. In Fig.~\ref{fig8} we plot the actual (blue circles) and approximate (red triangles) optimal synchronizability ratio $R_{\text{min}}$ vs. the root set fraction $n/N_1$ found for hierarchical products constructed using ER networks of size $N_1=50$ and $N_2=20$ using link probabilities $p=0.2$ and $0.5$, respectively. (These probabilities are chosen to attain a rough balance of the mean degree.) Results represent an average over $100$ networks, with standard deviation denoted by dashed curves. In general we see that the larger the root set fraction, the more synchronizable the hierarchical product can be when $\alpha$ is properly tuned. Thus, the more pathways built into the hierarchical product via the root set, the more favorable the synchronization properties.

\begin{figure}[t]
\centering
\epsfig{file =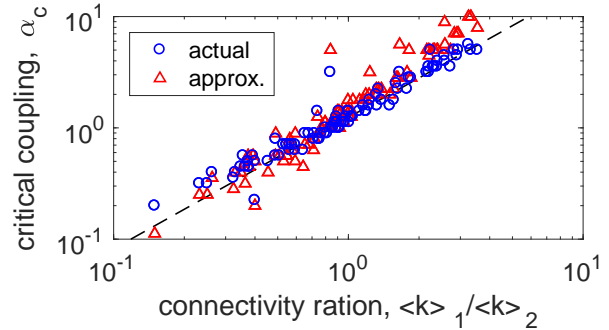, clip =,width=0.98\linewidth }
\caption{(Color online) {\it Critical coupling}. The actual (blue circles) and approximated (red triangles) critical coupling parameter $\alpha_c$ that optimized synchronizability ratio, $R=R_{\text{min}}$, as a function of the connectivity ratio $\langle k\rangle_1/\langle k\rangle_2$. 100 networks were constructed using $N_1=N_2=50$ with mean degrees determined randomly such that they fall between $5$ and $45$. Results fall roughly around the power-law relationship $\alpha_c\propto\left(\langle k\rangle_1/\langle k\rangle_2\right)^\beta$ (dashed black) for $\beta\approx1.18$.}
\label{fig9}
\end{figure}

A more interesting question, however, is at what coupling value $\alpha_c$ is the synchronizability of a hierarchical product optimized? We find that this critical coupling value does not depend significantly on the root set itself, but rather the contrast between the primary and secondary subnetworks. In fact, the answer to this question sheds light on role of the primary and secondary networks in relation to one another. We consider hierarchical products constructed from ER networks, both of size $N_1=N_2=50$, for each subnetwork choosing the link probabilities randomly to allow the mean degrees for the subnetworks, denoted $\langle k\rangle_1$ and $\langle k\rangle_2$, to vary between $5$ and $45$. Using the mean degree of each subnetwork as a proxy for overall connectivity, we then compare the critical coupling values $\alpha_c$ to the connectivity ratio $\langle k\rangle_1/\langle k\rangle_2$, plotting in Fig.~\ref{fig9} the results from $100$ different networks the actual (blue circles) and approximated (red triangles) results. Figure~\ref{fig9} shows a positive, roughly power-law relationship between $\alpha_c$ and $\langle k\rangle_1/\langle k\rangle_2$. This suggests that, to achieve optimal synchronizability, the coupling should be tuned to balance the connectivity properties of the primary and secondary subnetworks. If the ratio $\langle k\rangle_1/\langle k\rangle_2$ is large (i.e., larger than one), indicating that the primary subnetwork is more strongly connected than the secondary subnetwork, then the coupling should be increased to strengthen the secondary subnetwork in compensation. On the other hand, If the ratio $\langle k\rangle_1/\langle k\rangle_2$ is small (i.e., less than one), indicating that the primary subnetwork is connected more weakly than the secondary subnetworks, then the coupling should be decreased to weaken the secondary subnetwork in compensation. Moreover, we find that the results fall roughly around the power-law relationship $\alpha_c\propto\left(\langle k\rangle_1/\langle k\rangle_2\right)^\beta$ for $\beta\approx1.18$, as illustrated with the dashed black curve.

\section{Discussion}\label{sec5}

In this paper we have studied the long-term dynamics of diffusion and synchronization processes on hierarchical products. We have applied the methodology from previous work~\cite{Skardal2016PRE} characterizing the eigenvalues of the adjacency matrix of hierarchical products to the eigenvalues of the Laplacian matrix, allowing us to make analytical predictions for both diffusion and synchronization dynamics. In particular, this has allowed us to identify the roles that the primary and secondary subnetworks play in shaping the long-term dynamics in relation to a coupling parameter that weighs the contribution of the secondary subnetwork relative to the primary subnetwork. More generally our results explore the effects that different substructure of networks play in shaping large-scale dynamics by either promoting or inhibiting these processes. 

In the case of diffusion, we have identified two regimes corresponding to small and large coupling. In the small coupling regime the diffusion rate is slow, scaling with the coupling itself, and is completely determined by the structure of the secondary subnetwork. In the large coupling regime the diffusion rate saturates to a constant value which is determined by the structure of the primary subnetwork, as well as possibly the size of the root set and its structure. Thus, there is an transition that occurs as coupling is varied through intermediate values, both in terms of the long-term dynamical behavior, as well as the roles that the different structures that make up the hierarchical product play in shaping those dynamics.

In the case of synchronization, we find that the synchronization properties of hierarchical products are poor in both the small and large coupling regimes, but is optimized at an intermediate critical coupling value that minimizes the synchronizability ratio. In general, the optimal synchronizability ratio that a hierarchical product can attain, assuming $\alpha$ can be properly tuned, improves as the size of the root set increases. However, a more interesting phenomenon occurs with the critical coupling parameter that optimizes synchronization, which highlights the difference in overall connectivity between the primary and secondary subnetworks. Specifically, the critical coupling value is tuned to compensate for this difference, either strengthening or weakening the secondary subnetwork to bring is connectivity closer to that of the primary subnetwork.

Throughout this work we have focused on the case of undirected subnetworks, resulting in undirected hierarchical product. In the case of directed subnetworks it is straightforward to see that the resulting hierarchical product also becomes directed. In principle, the techniques used here to calculate eigenvalues may be preserved, however the emergence of complex eigenvalues may require some care when applying these results. This is also true when working with the asymmetric normalized Laplacian matrix for random walks, even in the case of undirected networks.

The class of networks investigated here, i.e., the hierarchical product~\cite{Barriere2009DAM1,Barriere2009DAM2}, represents a relatively wide subset possible generalizations of classical graph products~\cite{Hammack2011}. In general, graph products represent natural ways of building larger networks from two or more smaller subnetworks where the macroscopic properties of the large network can be understood in terms of the properties of the smaller subnetworks that comprise it. The general notion of a network consisting of smaller substructures remains a central theme in physics and mathematics, with examples including multilayer and multiplex networks~\cite{Granell2013PRL,Kouvaris2015SciRep,DeDomenico2016NatPhys,Nicosia2017PRL}, modular networks~\cite{Liu2005EPL,Skardal2012PRE}, hierarchical and hierarchical modular networks~\cite{Boettcher2008JPA,Boettcher2009PRE,Kaiser2010Front,Moretti2013NatComms}, and networks of networks~\cite{Gao2011PRL,Bianconi2014PRE}. Many of these cases share commonalities, for example the behavior of the Laplacian eigenvalues we observe in hierarchical products (e.g., see Figs.~\ref{fig2} and \ref{fig4}) reflects the behavior of Laplacian eigenvalues in multiplex networks~\cite{Gomez2013PRL,Sole2013PRE}. Given this overlap in phenomenological behavior, we hypothesize that understanding the macroscopic structural properties of hierarchical products is not only important in the context of graph products, but also more generally for wider classes of networks. To date, a handful of studies have investigated the overall structure of hierarchical products~\cite{Barriere2016JPA,Skardal2016PRE}, little work has focused on behavior of dynamical processes taking place on hierarchical products. 

Finally, we emphasize that the contributions of this paper, i.e., the description of the long-term diffusion and synchronization dynamics on hierarchical products, fit within the broader question of how various structures and organizations in complex networks dictate large-scale dynamical processes. Specifically, the findings presented here can be interpreted as investigating how different components and substructures of a given network, and their relative strengths, function in shaping the dynamics that occur across the whole network. This broad question has been investigated for various kinds of networks (e.g., see those listed above); here we study this broad question in the context of a graph product. In particular, the long-term behaviors of both diffusion and synchronization dynamics identify the role of the secondary subnetwork as a {\it connector} in comparison to the more central primary subnetwork, as well as the role that nodes in the root set play in facilitating these connections. Therefore, the coupling parameter modifies the relative strengths of the overall connectivity within the primary subnetworks compared to the connectivity between different subnetworks. Specifically, it is the secondary subnetwork structure that is responsible for the smaller eigenvalues in the small coupling regime, whereas in the large coupling regime the primary subnetwork is responsible for the smaller eigenvalues. More broadly, the transitions that we observe in the dynamics can be interpreted as a shift in which of these connectivities becomes the effective bottleneck for the dynamics and the primary hinderance for diffusion or spread of consensus (i.e., synchronization) throughout the network. 

\begin{appendix}

\section{Eigenvalue Perturbation Analysis}\label{app:A}

Here we present the perturbative analysis for the eigenvalues of the Laplacian $L_\alpha$ in Eq.~(\ref{eq:02}), which are in turn given by the eigenvalues of Eq.~(\ref{eq:03}). We consider here the $N_1(N_2-1)$ eigenvalues corresponding to inserting the nonzero eigenvalues $\mu_2,\dots,\mu_{N_2}$ into Eq.~(\ref{eq:03}). Beginning with the limit of small coupling, $\alpha\ll1$, we proceeding perturbatively as in Ref.~\cite{Skardal2016PRE}, we make the common notational change $\epsilon=\alpha$ such that $\epsilon\ll1$ is a small parameter and study the eigenvalues of
\begin{align}
L_\epsilon(\mu_i) = L_1+\epsilon\mu_i D_1.\label{eq:002}
\end{align}
In the limit $\epsilon\to0^+$ we recover the spectrum of $L_1$, i.e., eigenvalues $\nu_i$ and eigenvectors $\bm{v}^i$, and therefore we propose a perturbative ansatz of the form
\begin{align}
\lambda_j(\epsilon)&=\nu_j+\epsilon\hat{\lambda}_j + \mathcal{O}(\epsilon^2),\label{eq:003}\\
\bm{w}^j(\epsilon)&=\bm{v}^j+\epsilon\hat{\bm{w}}^j + \mathcal{O}(\epsilon^2),\label{eq:004}
\end{align}
and seek the coefficient $\hat{\lambda}_j$ of the first-order correction, i.e., searching for the leading order behavior of the Taylor series for $\lambda_j(\epsilon)$ and $\bm{w}^j(\epsilon)$. Inserting Eqs.~(\ref{eq:002}), (\ref{eq:003}), and (\ref{eq:004}) into the eigenvalue equation $L_\epsilon(\mu_i)\bm{w}^j(\epsilon)=\lambda_j(\epsilon)\bm{w}^j(\epsilon)$ and collecting the leading order terms at $\mathcal{O}(\epsilon)$, we obtain
\begin{align}
\mu_i D_1\bm{v}^j + L_1 \hat{\bm{w}}^j=\hat{\lambda}_j\bm{v}^j+\nu_j\hat{\bm{w}}^j.\label{eq:005}
\end{align}
Left-multiplying Eq.~(\ref{eq:005}) by $\bm{v}^{jT}$ and noting that the term on the left-hand side $\bm{v}^jL_1\hat{\bm{w}}^j=\nu_j\bm{v}^j\hat{\bm{w}}^j$ cancels with the right-hand side, we obtain
\begin{align}
\hat{\lambda}_j=\mu_i\bm{v}^{jT}D_1\bm{v}^j\label{eq:006}
\end{align}
We note that terms similar to the right hand side in Eq.~(\ref{eq:006}) appear often in perturbaitve analyses, and are akin to the first order correction to the energy of a Hamiltonian~\cite{Schrodinger}. Substituting back $\epsilon=\alpha$, we have that the eigenvalues of $L_\alpha(\mu_i)$ to leading order are given by
\begin{align}
\lambda_j(\alpha)=\nu_j+\alpha\mu_i\bm{v}^{jT}D_1\bm{v}^j,\label{eq:007}
\end{align}
giving the expression presented in Eq.~(\ref{eq:05}) in the main text.

Next we consider the limit of large coupling, $\alpha\gg1$, now letting $\epsilon=\alpha^{-1}$ be a small parameter. We again proceed perturbatively, noting that now
\begin{align}
\alpha^{-1}L_\epsilon(\mu_i)=\mu_i D_1 + \epsilon L_1,\label{eq:008}
\end{align}
so that after finding the eigenvalues of the right hand side of Eq.~(\ref{eq:008}) for $\mu_2,\dots,\mu_{N_2}$, we then multiply by $\alpha$ to obtain the final eigenvalues. As in Ref.~\cite{Skardal2016PRE}, the perturbative analysis for large coupling then becomes more complicated than that for small coupling due to the fact that when $\epsilon=0$ the right hand side of Eq.~(\ref{eq:008}) reduces to the matrix $\mu_i D_1$, which is degenerate. Specifically, $D_1$ has precisely $n$ eigenvalues equal to one and $(N_1-n)$ eigenvalues equal to zero, where $n$ is the number of nonzero entries of $D_1$, i.e., the size of the roots set $U$. We will refer to the eigenspaces associated with the one and zero eigenvalues of $D_1$ as the nontrivial and trivial eigenspaces. (Note that the trivial eigenspace is precisely the nullspace.) Specifically, the nontrivial eigenspace of $D_1$ is the span of all vectors whose entries are zero where the diagonal entries of $D_1$ are zero, and the trivial eigenspace of $D_1$ is the span of all vectors whose entries are zero where the diagonal entries of $D_1$ are non-zero. This requires us to consider two subcases of our asymptotic analysis: one for the nontrivial eigenspace which will yield $n$ eigenvalues and another for the trivial eigenspace which will yield $N_1-n$ eigenvalues.

We begin with the nontrivial eigenspace of $D_1$ and propose a perturbative ansatz of the form
\begin{align}
\tilde{\lambda}_j(\epsilon)&=\mu_i+\epsilon\hat{\lambda}_j + \mathcal{O}(\epsilon^2),\label{eq:009}\\
\bm{w}^j(\epsilon)&=\bm{x}+\epsilon\hat{\bm{w}}^j + \mathcal{O}(\epsilon^2),\label{eq:010}
\end{align}
where the vector $\bm{x}$ is in the non-trivial nullspace of $D_1$, i.e., $D_1\bm{x}=\bm{x}$. Inserting Eqs.~(\ref{eq:008}), (\ref{eq:009}), and (\ref{eq:010}) into the eigenvalue equation $\alpha^{-1}L_\epsilon(\mu_i)\bm{w}^j(\epsilon)=\tilde{\lambda}_j(\epsilon)\bm{w}^j(\epsilon)$ and collecting the leading order terms at $\mathcal{O}(\epsilon)$, we obtain
\begin{align}
\mu_i D_1\hat{\bm{w}}^j+L_1\bm{x}=\hat{\lambda}_j\bm{x}+\mu_i\hat{\bm{w}}^j.\label{eq:011}
\end{align}
Next, the entries of $\bm{x}$ that correspond to zeros in the diagonal of $D_1$ (i.e., nodes that do not belong to the root set $U$) are zero, so we eliminate these $(N_1-n)$ entries from Eq.~(\ref{eq:011}) and obtain the following $n$-dimensional vector equation:
\begin{align}
\begin{array}{rl}
\mu_i\hat{\bm{w}}^{\cancel{0}}+L_1^{\cancel{0}}\bm{x}^{\cancel{0}}&=\hat{\lambda}_j\bm{x}^{\cancel{0}}+\mu_i\hat{\bm{w}}^{\cancel{0}},\\
\to\hskip2exL_1^{\cancel{0}}\bm{x}^{\cancel{0}}&=\hat{\lambda}_j\bm{x}^{\cancel{0}},\end{array}\label{eq:012}
\end{align}
where $L_1^{\cancel{0}}$ is the $n\times n$ matrix obtained by keeping the rows and columns of $L_1$ corresponding to non-zero diagonal entries of $D_1$ and similarly $\hat{\bm{w}}^{j\cancel{0}}$ and $\bm{x}^{\cancel{0}}$ are the $n$-dimensional vectors obtained by keeping the same entries of $\hat{\bm{w}}^j$ and $\bm{x}$. Thus, $\hat{\lambda}_j$ is one of the $n$ eigenvalues of the matrix $L_1^{\cancel{0}}$, which we will denote $\nu_j^{\cancel{0}}$. Inserting this back into Eq.~(\ref{eq:009}), replacing $\epsilon=\alpha^{-1}$, and multiplying by $\alpha$, the $n$ eigenvalues of $L_\alpha(\mu_i)$ corresponding to the nontrivial eigenspace of $D_1$ to leading order are given by
\begin{align}
\lambda_j(\alpha)=\alpha\mu_i+\nu_j^{\cancel{0}}.\label{eq:013}
\end{align}

Turning our attention to the trivial eigenspace of $D_1$, we introduce a new perturbative anstaz:
\begin{align}
\tilde{\lambda}_j(\epsilon)&=0+\epsilon\hat{\lambda}_j + \mathcal{O}(\epsilon^2),\label{eq:014}\\
\bm{w}^j(\epsilon)&=\bm{y}+\epsilon\hat{\bm{w}}^j + \mathcal{O}(\epsilon^2),\label{eq:015}
\end{align}
where the vector $\bm{y}$ is now in the nullspace of $D_1$, i.e., $D_1\bm{y}=\bm{0}$. Inserting Eqs.~(\ref{eq:008}), (\ref{eq:014}), and (\ref{eq:015}) into the eigenvalue equation $\alpha^{-1}L_\epsilon(\mu_i)\bm{w}^j(\epsilon)=\tilde{\lambda}_j(\epsilon)\bm{w}^j(\epsilon)$ and collecting the leading order terms at $\mathcal{O}(\epsilon)$, we obtain
\begin{align}
\mu_i D_1\hat{\bm{w}}^j+L_1\bm{y}=\hat{\lambda}_j\bm{y}.\label{eq:016}
\end{align}
Similarly to Eq.~(\ref{eq:011}), several vector entries in Eq.~(\ref{eq:016}) are zero: this time all entries of $\bm{y}$ that correspond to ones in the diagonal of $D_1$ (i.e., nodes that are in the root set $U$) are zero. We therefore eliminate these $n$ entries from Eq.~(\ref{eq:016}) to obtain the following $(N_1-n)$-dimensional vector equation
\begin{align}
L_1^0\bm{y}^0=\hat{\lambda}_j\bm{y}^0,\label{eq:017}
\end{align}
where $L_1^0$ is the $(N_1-n)\times(N_1-n)$ matrix obtained by keeping the rows and columns of $L_1$ corresponding to zero diagonal entries of $D_1$ and similarly $\bm{y}^0$ is the $(N_1-n)$-dimensional vector obtained by keeping the same entries of $\bm{y}$. Thus, $\hat{\lambda}_j$ is an eigenvalue of the matrix $L_1^0$, which we will denote $\nu_j^0$. Inserting this back into Eq.~(\ref{eq:014}), replacing $\epsilon=\alpha^{-1}$, and multiplying by $\alpha$, the $(N_1-n)$ eigenvalues of $L_\alpha(\mu_i)$ corresponding to the trivial eigenspace of $D_1$ to leading order are given by
\begin{align}
\lambda_j(\alpha)=\nu_j^0.\label{eq:018}
\end{align}
Eqs.~(\ref{eq:013}) and (\ref{eq:018}) give those expressions presented in Eqs.~(\ref{eq:06}) and (\ref{eq:07}) in the main text.

\end{appendix}

\bibliographystyle{plain}

\end{document}